\documentclass[sigconf,natbib=true]{acmart}
\AtBeginDocument{%
  \providecommand\BibTeX{{%
    \normalfont B\kern-0.5em{\scshape i\kern-0.25em b}\kern-0.8em\TeX}}}

\usepackage{balance}
\usepackage{soul}
\usepackage{graphicx}
\usepackage{booktabs}
\usepackage{float}
\usepackage{overpic}
\usepackage{float}  
\usepackage{subfloat}
\usepackage{stfloats} 
\usepackage[skip=0.5\baselineskip]{caption}
\usepackage{bm}
\usepackage{amsmath}
\usepackage{booktabs}
\usepackage{multirow}
\usepackage{multicol}
\usepackage{enumitem}
\usepackage{subcaption}
\usepackage{threeparttable}
\usepackage{xspace}
\usepackage{color}
\usepackage[normalem]{ulem}
\usepackage{algorithmicx,algorithm}
\usepackage{algpseudocode}
\usepackage{marvosym}   
\usepackage{colortbl}   
\usepackage{tcolorbox}  
\tcbuselibrary{breakable}   
\usepackage{wasysym}    
\usepackage[figuresright]{rotating} 

\makeatletter
\def\@ACM@checkaffil{
    \if@ACM@instpresent\else
    \ClassWarningNoLine{\@classname}{No institution present for an affiliation}%
    \fi
    \if@ACM@citypresent\else
    \ClassWarningNoLine{\@classname}{No city present for an affiliation}%
    \fi
    \if@ACM@countrypresent\else
        \ClassWarningNoLine{\@classname}{No country present for an affiliation}%
    \fi
}
\makeatother

\usepackage{xspace}

\newcommand{\name}{GFlowGR\xspace}

\newcommand{\eat}[1]{}

\copyrightyear{2026}
\acmYear{2026}
\setcopyright{cc}
\setcctype{by-nc-nd}
\acmConference[SIGIR '26]{Proceedings of the 49th International ACM SIGIR Conference on Research and Development in Information Retrieval}{July 20--24, 2026}{Melbourne, VIC, Australia}
\acmBooktitle{Proceedings of the 49th International ACM SIGIR Conference on Research and Development in Information Retrieval (SIGIR '26), July 20--24, 2026, Melbourne, VIC, Australia}
\acmDOI{10.1145/3805712.3809663}
\acmISBN{979-8-4007-2599-9/2026/07}

\begin{document}
\begin{sloppypar}   
\title{GFlowGR: Fine-tuning Generative Recommendation Frameworks with Generative Flow Networks}

\author{Yejing Wang}
\email{yejing.wang@my.cityu.edu.hk}
\affiliation{
    \institution{City University of Hong Kong}
    \city{Hong Kong SAR}
    \country{China}
}
\authornote{Work was conducted during the internship of Yejing Wang at Alibaba.}
\authornote{Yejing Wang and Shengyu Zhou contributed equally to this work.}
\author{Shengyu Zhou}
\authornotemark[2]
\affiliation{
    \institution{Alibaba Group}
    \city{Beijing}
    \country{China}
}

\author{Jinyu Lu}
\affiliation{
    \institution{Alibaba Group}
    \city{Beijing}
    \country{China}
}

\author{Qidong Liu}
\affiliation{
    \institution{City University of Hong Kong}
    \city{Hong Kong SAR}
    \country{China}
}

\author{Xinhang Li}
\affiliation{
    \institution{Alibaba Group}
    \city{Beijing}
    \country{China}
}


\author{Wenlin Zhang}
\affiliation{
    \institution{City University of Hong Kong}
    \city{Hong Kong SAR}
    \country{China}
}

\author{Feng Li}
\affiliation{
    \institution{Alibaba Group}
    \city{Beijing}
    \country{China}
}

\author{Pengjie Wang}
\affiliation{
    \institution{Alibaba Group}
    \city{Beijing}
    \country{China}
}

\author{Chuan Yu}
\affiliation{
    \institution{Alibaba Group}
    \city{Beijing}
    \country{China}
}
\author{Jian Xu}
\affiliation{
    \institution{Alibaba Group}
    \city{Beijing}
    \country{China}
}

\author{Bo Zheng}
\authornotemark[3]
\email{bozheng@alibaba-inc.com}
\affiliation{
    \institution{Alibaba Group}
    \city{Beijing}
    \country{China}
}

\author{Xiangyu Zhao}
\email{xianzhao@cityu.edu.hk}
\affiliation{
    \institution{City University of Hong Kong}
    \city{Hong Kong SAR}
    \country{China}
}

\authornote{Bo Zheng and Xiangyu Zhao are the corresponding authors.}

\renewcommand{\shortauthors}{Yejing Wang et al.}
\begin{abstract}
Generative recommendation (GR) has shown great promise in industrial applications, particularly for candidate generation and end-to-end recommendations. However, existing GR training paradigms suffer from two fundamental mismatches with real-world deployment requirements. First, they optimize for \textbf{point-wise} prediction of a single ground-truth item, whereas practical systems must produce a diverse, high-value \textbf{set of candidates}. Second, they treat all user interactions as \textbf{equally informative}, ignoring their inherent \textbf{differences in utility}. 
Although reward-based fine-tuning offers a partial remedy, it often lacks token-level supervision. 
To address these challenges, we reformulate GR as a sequential set-generation problem and propose \name, a GFlowNet-based fine-tuning framework that explicitly aligns generation probabilities with item-level utilities.
\name comprises three tightly integrated components, each addressing a key limitation of conventional fine-tuning: a \textbf{trajectory sampler} that constructs training trajectories from candidate sets to enable \textit{set-wise} learning, a behavior-aware \textbf{reward model} that quantifies item utility to support \textit{value-aware} optimization, and a \textbf{ GFlowNet objective} that provides \textit{token-level} supervision.
Extensive experiments on three real-world datasets with two representative LLM-based GR backbones show consistent and significant improvements over strong baselines, validating the effectiveness of our approach.
For real-world deployment, \name has been integrated into \textbf{Taobao}'s search advertising businesses, delivering a \textbf{0.4\%} relative improvement in annual revenue since its launch in mid-2025,  corresponding to \textbf{billion-level} monetary gains.
Code is available at \url{https://github.com/Applied-Machine-Learning-Lab/SIGIR26_GFlowGR}.
\end{abstract}





\maketitle

\section{Introduction}
Large Language Models (LLMs) have demonstrated exceptional capabilities in understanding, reasoning, and sequence generation, leading to their widespread adoption across diverse domains~\cite{naveed2023comprehensive,liu2025large,gao2025llm4rerank,wang2024llm4msr,liu2024large,zhang2025notellm,gao2025generative,zhang2025llm}. 
In information retrieval and recommendation systems~\cite{wang2022autofield,wang2023plate}, large language models (LLMs) are increasingly used to power next-generation recommender systems~\cite{liu2025best,liu2024llm,liu2025bridge,zhang2025notellm,wang2025rethinking,fu2025unified}, such as generative recommendation (GR) frameworks~\cite{rajput2023recommender,huang2024improving,hua2023index}. These frameworks have seen growing adoption in large-scale industrial services, such as e-commerce and online advertising~\cite{zhou2025onerec,zheng2025ega,guo2025onesug,wei2025oneloc,wang2025nezha}, where they are deployed for candidate generation or for end-to-end recommendations.

\begin{figure*}[t]
\centering
\includegraphics[width=1\textwidth]{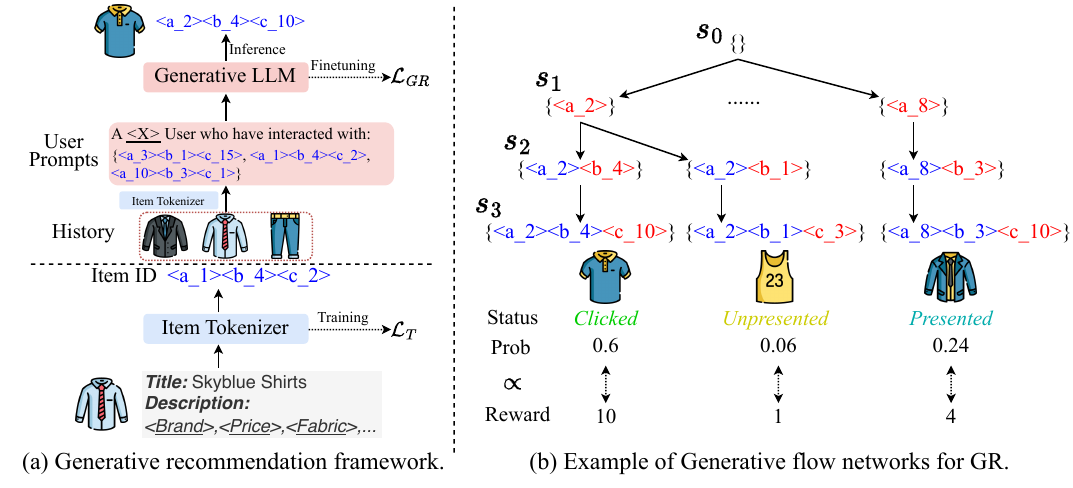}
\caption{Overview of GR and GFlowNets, with the length of item identifiers $L=3$. (a) A process for constructing the GR framework. (b) Generative flow structure is exemplified with a set of user interactions, which typically includes positive feedback (``Clicked''), impression (``Presented''), and unexposed items (``unpresented'').}
\label{fig:pre}
\end{figure*}

As visualized in Figure~\ref{fig:pre} (a), general GR frameworks typically consist of three key components~\cite{li2024survey,li2025survey}: an \textit{item tokenizer} that maps items into discrete token sequences, a \textit{user prompt} that contextualizes historical interactions, and a \textit{generative LLM} that autoregressively produces recommendation tokens~\cite{li2024survey}. While significant progress has been made in item tokenization~\cite{zheng2025universal,liu2024end,lin2025order,wang2024learnable,zhu2024cost}, prompting strategies, and LLM inference efficiency~\cite{lin2025efficient,ding2024inductive,wang2025nezha}, a critical bottleneck remains in the \textbf{training paradigm of LLMs}—particularly its misalignment with real-world deployment.

The prevailing approach, Supervised Fine-Tuning (SFT)~\cite{rajput2023recommender}, optimizes the model to reproduce a single ground-truth item per training instance. 
This formulation is fundamentally at odds with production requirements from two perspectives:
First, SFT ignores the full set of relevant items during training, optimizing only for a single target, despite the fact that deployment demands set-wise generation. This creates a persistent gap between offline training that focuses on single-item prediction and online inference, where generating multiple high-quality candidates is essential.
Second, SFT treats all user interactions (e.g., impressions, clicks, purchases) as equally valuable, disregarding their varying business or user-intent significance. For example, a purchase signal indicates stronger intent and higher platform value than a mere impression, and should therefore be assigned a higher generation probability. 
By overlooking such distinctions, SFT yields suboptimal ranking and calibration in production environments.



To mitigate this, recent work has explored reward-based fine-tuning to incorporate value-aware signals and consider multiple samples~\cite{chen2025onesearch,zhou2025onerec}. 
Consequently, recent efforts have explored reinforcement learning (RL) fine-tuning strategies~\cite{zhao2017deep,zhao2018recommendations} that incorporate value-aware signals as reward and utilize multiple interaction samples to better align training with deployment objectives.
For instance, \citet{chen2025onesearch} adapt Direct Preference Optimization (DPO)~\cite{rafailov2023directDPO} by constructing preference pairs from interaction histories, while OneRec-V2~\cite{deng2025onerec} extends group-wise policy optimization (GRPO)~\cite{shao2024deepseekmathGRPO} to leverage full interaction sets. However, these methods assign rewards only at the item level and lack supervision over the underlying token-level generation process. Since items in GR are represented as multi-token sequences (Figure~\ref{fig:pre}(b)), this absence of fine-grained, token-level feedback limits both optimization precision and generation quality~\cite{gao2025process,guo2026promise}. \textit{Thus, an ideal fine-tuning framework must jointly support \textbf{(i) learning from value-differentiated item sets} and \textbf{(ii) providing direct token-level supervision}.}

Generative Flow Networks (GFlowNets) offer a compelling foundation for addressing this challenge. By modeling generation as a stochastic policy over discrete trajectories and assigning terminal rewards based on item utility, GFlowNets naturally decompose item-level rewards into flow-consistent token-level targets~\cite{bengio2023gflownet}. Crucially, they enforce that the probability of generating any complete item sequence is proportional to its reward, thereby promoting value alignment. Recent studies have successfully applied GFlowNets to multi-step reasoning~\cite{kang2025gflowvlm,takase2024gflownet,yu2024flowfor} and compositional generation tasks~\cite{jain2022biological}. Despite promising results in traditional recommendation settings~\cite{liu2023generative,liu2024modeling} and LLM-based applications~\cite{gao2025process}, a principled GFlowNet framework tailored for GR remains unexplored.

To fill this gap, we propose \name, a GFlowNet-based fine-tuning paradigm specifically designed for GR. \name reformulates GR as a sequential set-generation problem and introduces three tightly integrated components: (1) a trajectory sampler that constructs training trajectories from candidate sets, (2) a behavior-aware reward model that quantifies item utility from heterogeneous user signals, and (3) a GFlowNet objective that provides token-level learning signals while ensuring the model’s output distribution aligns with the reward landscape. The framework is modular, compatible with existing GR backbones, and serves as a drop-in enhancement to standard SFT.
Key contributions are:
\begin{itemize}[leftmargin=*]
\item We formalize GR as a multi-step, value-aware set generation task and propose \name, the first GFlowNet-based fine-tuning framework for GR that enables token-level supervision.
\item We present a modular design for \name, comprising a trajectory sampler, a reward model, and a GFlowNet loss, and demonstrate its flexibility through multiple practical instantiations.
\item We conduct extensive experiments across three real-world datasets using two representative LLM-based GR backbones, demonstrating consistent and significant improvements over strong baselines.
\item We report successful large-scale deployment of \name in \textbf{ Taobao's search advertising businesses}, serving \textbf{hundreds of millions} of daily active users and delivering a \textbf{0.4\% relative increase} in annual revenue, resulting in \textbf{billion-level} daily monetary gains at scale.
\end{itemize}


\section{Preliminary}

\textbf{Generative Recommendation (GR). }
In this paper, we define GR frameworks by referring to recent work \cite{rajput2023recommender,wang2024learnable}. A typical GR workflow is illustrated in Figure~\ref{fig:pre} (a). The framework first constructs an item tokenizer using item features to decompose items $v$ into token sequences $\{t_l\}_{l<L}$, serving as item identifiers. The lower portion of the figure exemplifies that a skyblue shirt is encoded as the token sequence $<a\_1><b\_4><c\_2>$ based on its title and descriptions. To infer user preferences, GR frameworks assemble user prompts (LLM inputs) by integrating task descriptions, historical user interactions, and profile information (e.g., age and gender, denoted as $<\underline{X}>$). These prompts, exemplified by tokenized interaction histories (blue tokens for jackets, shirts, pants) and profiles, are fed into LLMs to generate token sequences representing the most probable next item. As shown in the upper portion of Figure~\ref{fig:pre} (a), this process produces $<a\_2><b\_4><c\_10>$, decoding to a blue polo shirt as the recommended item.

To achieve the recommendation goal, the GR training paradigm comprises two stages: optimizing the item tokenizer ($\mathcal{L}_T$) and the generative LLMs ($\mathcal{L}_{\text{GR}}$). For the tokenizer, one prevalent tokenization strategy is based on RQ-VAE solutions~\cite{lee2022autoregressive,rajput2023recommender}, where $\mathcal{L}_T$ includes a codebook loss and a reconstruction loss. Studies enhancing the tokenizer often incorporate auxiliary losses into $\mathcal{L}_T$, such as diversity~\cite{wang2024learnable} and collaborative-constrained losses~\cite{lin2025order}. 
$\mathcal{L}_{\text{GR}}$ typically adopts the next-token prediction loss~\cite{deng2025onerec,rajput2023recommender,wang2024learnable,liu2024multi}$:$  
\begin{gather}
    \mathcal{L}_{\text{GR}}(U,v)=-\sum_{l=1}^L \mathrm{log}\mathbb{P}_{\text{GR}}(t_l|U,\{t_i\}_{i<l})\label{eq:gr_loss}
\end{gather}
where $L$ denotes the length of item identifiers (e.g., $L = 3$ for the visualized examples in this paper). We denote the probability that LLMs generate the next ground-truth token $t_l$ given the prefix $U, \{t_i\}_{i<l}$ as $\mathbb{P}\left(t_l \,\big|\, U, \{t_i\}_{i<l}\right)$. Here, $\{t_i\}_{i<l}$ represents the item identifier tokens preceding $t_l$.

\noindent\textbf{Generative Flow Networks.}
The core idea of Generative Flow Networks (GFlowNets) stems from the generation of compositional objects through a sequence of stochastic steps. We formally define the state transition sequence as $\tau = (s_0 \to s_1 \to \dots \to s_L)$, where $s_L$ represents the terminal generation state and $\tau$ denotes a specific generation trajectory. The training objective of GFlowNets is to align the generation probability of any trajectory $\tau$ with the reward obtained at its terminal state, such that $\mathbb{P}(\tau) \propto \mathcal{R}(s_L)$ ~\cite{bengio2021flow,takase2024gflownet}. To achieve this, GFlowNets introduce a flow estimator $\mathcal{F}(s_l)$ to model the likelihood of traversing state $s_l$, transforming the proportionality objective into a flow-matching principle: the incoming flow to a state must equal the outgoing flow. This is formalized as:  
\begin{gather}
\mathcal{F}(s_l)\mathbb{P}_F(s_{l+1}|s_l)=\mathcal{F}(s_{l+1})\mathbb{P}_B(s_{l}|s_{l+1})
\end{gather}
Here, $\mathbb{P}_F(s_{l+1} | s_l)$ denotes the forward transition probability from state $s_l$ to $s_{l+1}$, while $\mathbb{P}_B(s_l | s_{l+1})$ represents the backward probability that state $s_{l+1}$ originated from $s_l$. This bidirectional flow balance ensures that trajectories with higher terminal rewards $\mathcal{R}(s_L)$ are assigned greater generation probabilities, enabling efficient exploration of complex combinatorial spaces~\cite{bengio2023gflownet}.

To facilitate practical implementation, two optimization-friendly variants are derived from the flow-matching principle~\cite{malkin2022trajectory,bengio2021flow}: detailed balance (DB) and trajectory balance (TB). The former focuses on stepwise flow equilibrium, while the latter enforces balance at the full-trajectory level. Their respective loss functions are defined:  
\begin{gather}
\mathcal{L}_{\text{DB}}(\tau)=\sum_{l=0}^{L-1}\Big(\mathrm{log}\frac{\mathcal{F}(s_l)\mathbb{P}_F(s_{l+1}|s_l)}{\mathcal{F}(s_{l+1})\mathbb{P}_B(s_{l}|s_{l+1})}\Big)^2 \label{eq:dbloss_def} \\
\mathcal{L}_{\text{TB}}(\tau)=\Big(\mathrm{log}\frac{\mathcal{Z}\prod_{l=0}^{L-1}\mathbb{P}_F(s_{l+1}|s_l)}{\mathcal{R}(s_{L})\prod_{l=0}^{L-1}\mathbb{P}_B(s_{l}|s_{l+1})}\Big)^2 \label{eq:tbloss_def} 
\end{gather} 
Here, $\mathcal{Z}$ is a global scalar that approximates $\mathcal{F}(s_0)$~\cite{malkin2022trajectory}. 

\begin{figure*}[t]
\centering
\includegraphics[width=0.95\textwidth]{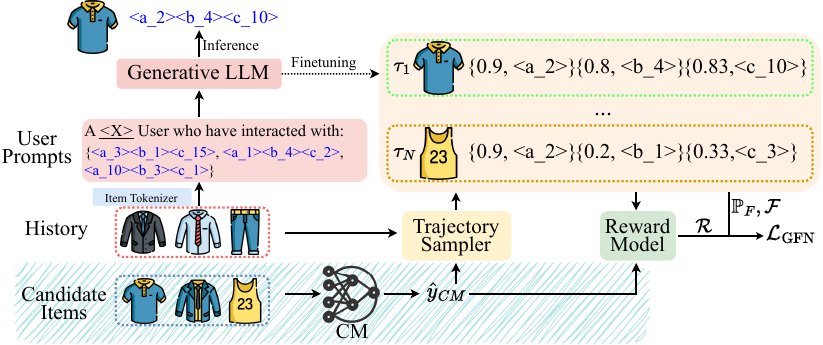}
\caption{Framework of \name. The visualized example shows a the learning on a $N$-item set illustrated with two items among the set, including a positive sample (blue polo shirt) and an unpresented sample (yellow undershirt). Forward probabilities ($\mathbb{P}_F$) for both trajectories are displayed (e.g., $0.9, 0.8, 0.83$ for $\tau_1$ ), while estimated flow values ($\mathcal{F}$) are omitted for brevity. The bluesky shaded area presents a example of incorporating collaborative models (CM).}
\label{fig:frame}
\end{figure*}

\noindent\textbf{Motivation.} 
Two key insights motivate the application of GFlowNets to GR: 
\textbf{First, GR is inherently a multi-step sequential generation process}, which aligns naturally with the trajectory-based formulation of GFlowNets. 
As shown in Figure~\ref{fig:pre} (b), generating the next item identifier unfolds over $L$ sequential steps: starting from an initial state $s_0 = \{\emptyset\}$, LLMs iteratively generate tokens (e.g., $<a\_2>$, $<b\_4>$, $<c\_10>$) to form intermediate states $s_1 = \{<a\_2>\}$, $s_2 = \{<a\_2><b\_4>\}$, and finally reach the terminal state $s_3 = \{<a\_2><b\_4><c\_10>\}$, which decodes to a specific item (e.g., a blue polo shirt). This stepwise construction of item identifier fits precisely into the GFlowNet framework, where each complete sequence corresponds to a generation trajectory $\tau = (s_0 \to s_1 \to \dots \to s_L)$.

\textbf{Second, GFlowNets provide two critical capabilities that directly address the core limitations of existing paradigms:}

\textbf{\underline{(a) Full Set Utilization}}: 
Unlike Supervised Fine-Tuning (SFT), which conditions training on a single positive item, GFlowNets operate over sets of trajectories derived from all observed user interactions—including clicks, impressions, and even negative samples. By learning from this richer interaction set within a unified training objective, GFlowNets effectively bridge the gap between point-wise offline training and set-wise online serving.

\textbf{\underline{(b) Item Value Perception}}:
GFlowNets enforce a fundamental relationship between generation probability and item reward: $\mathbb{P}(\tau) \propto \mathcal{R}(s_L)$. This ensures that high-reward (i.e., high-value) items are assigned higher generation probabilities—a property rigorously established in theory~\cite{bengio2021flow,bengio2023gflownet} and validated in diverse domains such as DNA design~\cite{jain2022biological} and multi-step reasoning~\cite{kang2025gflowvlm,yu2024flowfor}. In recommendation, this translates to proper ranking: preferred items are more likely to be generated than low-value ones.

As illustrated in the lower part of Figure~\ref{fig:pre}(b), by associating generation probabilities with estimated rewards and incorporating all types of user interactions—including clicked items, presented-but-not-clicked items, and even unpresented negatives—GFlowNets enable generative recommendation to prioritize high-value candidates. For instance, the blue jacket (which was presented and likely to be clicked) receives a higher reward and consequently a higher generation probability, whereas the yellow undershirt—an unpresented item with lower inferred utility—is assigned a lower probability, reflecting its reduced relevance.

From an implementation perspective, adapting GFlowNets to GR is conceptually straightforward. The forward transition probability $\mathbb{P}_F(s_{l+1}|s_l)$ directly corresponds to the LLM's token-level generation probability $\mathbb{P}_{\text{GR}}(t_{l+1}|U, s_l)$. Meanwhile, the backward probability $\mathbb{P}_B(s_l|s_{l+1}) = 1$, since each state $s_{l+1}=\{s_l,t_{l+1}\}$, $s_{l+1}$ is uniquely determined by its predecessor due to the deterministic token concatenation process. Consequently, the primary design challenges reduce to two components: (1) constructing meaningful training trajectories $\tau$ from user interaction sets, and (2) designing a behavior-aware reward model $\mathcal{R}(s_L)$ that accurately reflects item utility. These components form the backbone of our proposed framework and are detailed in the following sections.

\section{Methodology}
In this section, we detail the proposed \name framework, including the organization of trajectory sampler, reward model, and GFN losses. We also present the possible integration of existing collaborative models and elaborate the training and inference procedures. 

The architecture of \name is depicted within Figure~\ref{fig:frame}. Notably, \name modifies only the LLM fine-tuning stage for GR while keeping the inference pipeline unchanged.
The overall workflow is as follows: for a given input pair $(U,v)$, where $v$ denotes the single item with the positive feedback\footnote{In fact, the number of positive items is usually larger than one in the session. We discuss with the single positive situation for clarity.}.  Trajectory sampler selects augmented samples or loads the collected system logs to extend the item set as $\{v_n\}_{n\leq N})$, where $v_1=v$. GFlowNet then estimates the forward probabilities with $\mathbb{P}_F$ and flow values with $\mathcal{F}$ for all trajectories $\tau_n$ (corresponding to $v_n$). 
The reward model then assigns a reward $\mathcal{R}$ for each trajectory. Finally, \name computes the GFlowNet loss $\mathcal{L}_{\text{GFN}}$ for the LLM fine-tuning.

It is also possible for the framework to integrate a existing collaborative model (CM): for the candidate item $v_n$, CM is capable of providing the score based on user interaction histories for both the trajectory sampler and reward model.


\subsection{Trajectory Sampler}\label{sec:sampl} 

This section details the construction of the trajectory sampler, which produces the learning trajectories for a given user-interaction pair $(U,v)$. There are four methods for constructing the trajectory set: interaction log, random sampling, CM-based sampling, and on-policy sampling. 

\noindent\textbf{1. Interaction log. } For the industrial system, it is straightforward to cutoff the user interaction log for the trajectory set, which typically contains more than $N$ samples. Specifically, with this solution, the trajectory sampler simply finds the session for $(U,v)$ and selects the $N-1$ samples to formulate the item set.

\noindent\textbf{2. Random sampling. } The sparsity of recommendation data is a fundamental challenge, as only a minimal fraction of user-item interactions are observed~\cite{gao2022kuairec}. Consequently, constructing training trajectories solely from collected data is sometimes insufficient for \name, rendering the trajectory sampler essential to augment the sparse dataset. Consequently, we can also conduct random negative sampling~\cite{ma2024negative} for the trajectory sampler. For instance, randomly selecting $N-1$ augmented samples $\{v_n\}_{2\leq n\leq N}$ from the item set such that $v_m \neq v_n$ for all $1 \leq m,n\leq N,m\neq n$~\cite{yang2020mixed,wang2021cross}.

However, such randomness introduces training instability and often yields samples misaligned with the model’s evolving learning dynamics. For example, an obviously irrelevant augmented sample that was effective as a negative in early training epochs may become uninformative in later stages, as the model learns to easily distinguish it from true positives.

To address this, we propose two adaptive trajectory samplers that delivers high-quality augmented samples to \name in a curriculum-style manner. We assume that during the early training phase, LLMs struggle to distinguish between positive and augmented samples, while in later stages, overly simple samples contribute minimally to model improvement~\cite{wang2021survey}. 

\noindent\textbf{3. CM-based sampling. } Inspired by recent work demonstrating the benefits of integrating collaborative knowledge into LLM-based recommendation systems~\cite{lin2025can}, we leverage a pretrained collaborative model (CM) to guide sample augmentation. Specifically, we score $2N-1$ candidate augmentations $\{v_n\}_{2\leq n\leq 2N}$ using $\hat y_{CM}^{v_n} = \mathrm{CM}(v_n, U)$, where $\mathrm{CM}(\cdot)$ denotes the collaborative model and $\hat y_{CM}^{v_n}$ represents the collaborative score of item $v_n$ to user $U$. During initial training epochs, we prioritize sampling $N-1$ items with the lowest $\hat y_{CM}^{v_n}$ (easy negatives), such as the yellow undershirt in Figure~\ref{fig:pre}(b) and Figure~\ref{fig:frame}, which has low relevance to a user preferring blue formal attire. As training progresses, the sampler gradually incorporates harder augmented samples with higher collaborative scores—for example, the blue jacket in the visualized example, which aligns more closely with the user’s preferences but remains unobserved in historical data.

\noindent\textbf{4. LLM-based sampling. } An instant alternative to CM-based sampling is the on-policy sampling~\cite{deng2025onerec,gao2025sprec}. Specifically, the trajectory sampler can feed the user prompt to the current LLM to generate top-$(N-1)$ negatives for each $U$ with beam search.

These curriculum-based approaches enable the sampler to adaptively select trajectories, reducing randomness and enhancing training stability by ensuring augmented samples remain challenging yet relevant at each stage of learning.

\let\oldnl\nl
\newcommand{\nonl}{\renewcommand{\nl}{\let\nl\oldnl}}
\begin{algorithm}[t]
\caption{Train and inference process of \name with CM.} \label{alg:train}
\raggedright
\begin{algorithmic} [1]
    \State Indicate the tokenizer.
    \State Indicate the LLMs.
    \State Indicate the collaborative model ($\mathrm{CM}$).
\end{algorithmic}

\textbf{Train Process} 
\setcounter{algorithm}{3}
\begin{algorithmic} [1]
    \makeatletter
    \setcounter{ALG@line}{4}
    \State Train the tokenzier with $\mathcal{L}_T$ and represent items as identifiers with the trained tokenizer.
    \State Indicate the hyperparameters for the training, including the number of augmented trajectories ($N$), the weight of GFlowNets loss ($\lambda$).
    \For {a batch of samples $(U,v)$}
        \State Construct the user prompt with item tokenizer
        \State Get the trajectory $\tau_1$ for the ground-truth $v$.
        \State Generate $N-1$ augmented trajectories $\{\tau_n\}_{2\leq n\leq N}$ with the trajectory sampler with one of the strategy described in Section~\ref{sec:sampl}.
        \State Calculate the step-wise forward probability $\mathbb{P}_{\text{GR}}(t_{l+1}|U, s_l)$ of LLM for all trajectories. 
        \State Calculate the step-wise flow $\mathcal{F}(s_l)$ with the flow estimator. 
        \State Calculate the reward $\mathcal{R}(s_{L})$ for the terminal state.
        \State Calculate $\mathcal{L}_{\text{GFN}}$ with Equation (5) or Equation (6). 
        \State Calculate the SFT loss $\mathcal{L}_{\text{GR}}$ based on Equation (1).
        \State Sum the SFT loss and GFlowNets loss for the final loss $\mathcal{L}$ as in Equation~(7).
        \State Update LLM parameters.
    \EndFor
\end{algorithmic}

\textbf{Inference Process}
\setcounter{algorithm}{12}
\begin{algorithmic} [1]
    \makeatletter
    \setcounter{ALG@line}{13}
    \State Load the LLM checkpoint and item tokenizer.
    \For {arbitary user $\mathcal{U}$}
    \State Construct the user prompt with item tokenizer.
        \State Generate $L$ tokens as the identifier of the next item.
        \State Decode the identifier and return the predicted item.
    \EndFor
\end{algorithmic}
\end{algorithm}

\subsection{Reward Model}\label{sec:rew} 
The reward model is pivotal in GFlowNets, as it directly aligns generation probabilities with task-specific objectives \cite{bengio2021flow,bengio2023gflownet}. In recommendation contexts, prior GFlowNet reward models typically rely on observed user interactions (e.g., item ratings from 1–6 \cite{liu2024modeling,liu2023generative}), but this approach fails for unobserved samples. Assigning zero rewards to augmented samples may ignore the latent relevance and exacerbate exposure bias~\cite{chen2023bias}.  


To address this, we introduce a comprehensive reward model that can integrate multiple reward signals.
For simplicity, we describe the framework using a single augmented sample ($N=2$), which can be easily extended to multiple samples. Let $\tau = \{s_l\}_{l \leq L}$ denote the trajectory of the observed positive item $v$, and $\tau' = \{s'_l\}_{l \leq L}$ denote the trajectory of an augmented item $v'$. Several aspects of reward might be defined as:
\begin{itemize}[leftmargin=*]
    \item Interaction Signal ($r_a$): marks the interaction level for items.  For example, $r_a = 10$ for the liked item trajectory $\tau$, $r_a = 1$ for the clicked item trajectory $\tau$, and $r'_a = 0$ for the augmented or unpresented trajectory $\tau'$.
    \item  Estimated Score ($r_c$): the collaborative score from CM, i.e., $r_c = \hat{y}_{CM}^v, r'_c = \hat{y}_{CM}^{v'}$ or the LLM score score (forward probability).  
    \item Business Target: customized reward for different business targets~\cite{zhou2025onerec}. We illustrate with the format reward ($r_{\text{sim}}$), which reveals partial correctness by measuring the number of shared tokens with the positive sample. Specifically, for $\tau'$, $r'_{\text{sim}} =\sum_{l=1}^L \mathbb{I}(t_l = t'_l)$, where $\mathbb{I}(\cdot)$ is an indicator function. And $r_{\text{sim}}=L$ for $\tau$. For the visualized example, $ r'_{\text{sim}}=1$, as it has one shared token with the ground truth.
\end{itemize}

\subsection{GFN Training Objective}
The training objective of \name integrates two complementary loss components: the vanilla next-token prediction loss $\mathcal{L}_{\text{GR}}$ for positive samples, the GFlowNet training loss $\mathcal{L}_{\text{GFN}}$ over sampled trajectories. The inclusion of $\mathcal{L}_{\text{GR}}$ (Equation~\eqref{eq:gr_loss}) ensures the model retains foundational recommendation capabilities while adapting the LLM to GR tasks. For $\mathcal{L}_{\text{GFN}}$, substituting the forward probability $\mathbb{P}_F(s_{l+1}|s_l) = \mathbb{P}_{\text{GR}}(t_{l+1}|U, s_l)$ and backward probability $\mathbb{P}_B(s_{l}|s_{l+1}) = 1$ into Equation~\eqref{eq:dbloss_def} and \eqref{eq:tbloss_def} yields:
\begin{gather}
\mathcal{L}_{\text{DB}}(\tau)=\sum_{l=0}^{L-1}\Big(\mathrm{log}\frac{\mathcal{F}(s_l)\mathbb{P}_{\text{GR}}(t_{l+1}|U,s_l)}{\mathcal{F}(s_{l+1})}\Big)^2 \label{eq:dbloss_final} \\
\mathcal{L}_{\text{TB}}(\tau)=\Big(\mathrm{log}\frac{\mathcal{Z}\prod_{l=0}^{L-1}\mathbb{P}_{\text{GR}}(t_{l+1}|U,s_l)}{\mathcal{R}(s_{L})}\Big)^2 \label{eq:tbloss_final} 
\end{gather}
$\mathcal{F}$ and $\mathcal{Z}$ are learnable GFlowNet components, with $\mathcal{F}(s_L) = \mathcal{R}(s_L)$ in Equation~\eqref{eq:dbloss_final}. $\mathcal{L}_{\text{GFN}}$ can be instantiated using either $\mathcal{L}_{\text{DB}}$ or $\mathcal{L}_{\text{TB}}$. 
The complete loss function for $(U, v, \{\tau_n\}_{n\leq N})$ is:
\begin{gather}
  \mathcal{L}(U, v, \{\tau_n\}_{n\leq N}) = \mathcal{L}_{\text{GR}}(U, v) +\lambda \sum_{n=1}^N  \mathcal{L}_{\text{GFN}}(\tau_n) \label{eq:final_loss}
\end{gather}
where $\lambda$ is the hyperparameter to balance different objectives. 


\subsection{Training and Inference Process}
We delineate the training and inference procedures of \name in Algorithm~\ref{alg:train}, with the use of CM. Notably, \name can still work without CM.  Initially, the employed item tokenizer, generative LLMs for GR frameworks, and the collaborative model for \name are declared (lines 1 - 3). Subsequently, we train the item tokenizer and utilize the trained tokenizer to transform items into identifiers consisting of $ L$ tokens (line 5). Prior to fine-tuning LLMs with \name, we declare the essential hyperparameters (line 6). For a batch of training samples, we first structure the LLM input (line 8) and convert the item $v$ into the generation trajectory $\tau$ (line 9). Next, \name generates several augmented trajectories (line 10) and computes GFlowNets elements, which include the forward probability, flow, and reward (lines 11 - 13). Finally, the LLM parameters are updated by the sum of the SFT loss and the GFlowNets loss (lines 14 - 17). During the inference phase, we structure the LLM input, generate the subsequent tokens with the checkpoint, and decode the outcome (lines 16 - 18), where \name is not involved.

\section{Experiments}
To comprehensively evaluate the effectiveness and robustness of \name, we conduct extensive experiments on three real-world datasets using two representative LLM-based generative recommendation backbones. Our experimental design is guided by the following research questions:

\begin{table}[t]
\centering
\caption{The statistics of the preprocessed datasets}
\begin{tabular}{ccccc}
\toprule[1pt]
\textbf{Dataset} & \textbf{\# Users} & \textbf{\# Items} & \textbf{Sparsity} & \textbf{Avg.length} \\ 
\midrule
Yelp & 30,431 & 20,032 & 99.85\% & 9.39 \\
Beauty & 22,362 & 12,101 & 99.92\% & 7.87 \\
Instruments & 24,772 & 9,922 & 99.99\% & 7.32 \\ 
\bottomrule[1pt]
\end{tabular}
\label{tab:exp_dataset}
\end{table}

\begin{itemize}[leftmargin=*]
    \item \textbf{(RQ1)}: How does \name compare against current training paradigms in terms of recommendation accuracy?
\item \textbf{(RQ2)}: How sensitive is \name to critical hyper-parameters?
\item\textbf{(RQ3)}: What is the individual contribution of the trajectory sampler and the components of the reward model?
\item\textbf{(RQ4)}: How do different mechanisms for combining reward signals affect performance?
\item\textbf{(RQ5)}: How can LLMs further benefit \name?
\item\textbf{(RQ6)}: Can we qualitatively observe that \name generates more value-aligned and diverse recommendations?
\end{itemize}


\begin{table*}[t]
\tabcolsep=0.13cm 
\centering
\caption{Overall performance of baselines and \name. The boldface refers to beating all baselines. ``\textbf{{\Large *}}'' indicates the statistically significant improvements (i.e., one-sided t-test with $p<0.05$) over the best baseline. For all metrics, the higher is better.}
\label{tab:ovall}
\resizebox{\textwidth}{!}{%
\begin{tabular}{@{}c|c|cccc|cccc|cccc@{}}
\toprule
 \multirow{2}{*}{Model} & \multirow{2}{*}{Finetuning}                 & \multicolumn{4}{c|}{Beauty} & \multicolumn{4}{c|}{Instruments}& \multicolumn{4}{c}{Yelp} \\ 
&  & R@5  & R@10  & N@5  & N@10  & R@5  & R@10 & N@5 & N@10 & R@5  & R@10 & N@5 & N@10 \\ \midrule
\multirow{9}{*}{TIGER}                 
& SFT       &   0.0354   &  0.0558 &0.0234&0.0299&0.0864&0.1071&0.0740&0.0806
&0.0204&0.0334&0.0132&0.0174\\
& GRPO      &0.0347&0.0543&0.0229&0.0291&0.0834&0.0988&0.0718&0.0768&0.0193&0.0328&0.0126&0.0169\\
& DPO       &0.0308&0.0483&0.0200&0.0256&0.0803&0.0977&0.0692&0.0753&0.0100&0.0182&0.0060&0.0087\\
& S-DPO      &0.0314&0.0492&0.0208&0.0265&0.0791&0.0930&0.0688&0.0733&0.0212&0.0333&0.0137&0.0176\\
& SPRec      &0.0312&0.0487&0.0205&0.0258&0.0773&0.0903&0.0662&0.0709&0.0103&0.0186&0.0067&0.0090\\
& IPA      &0.0306&0.0481&0.0198&0.0245&0.0769&0.0899&0.0661&0.0707&0.0101&0.0179&0.0066&0.0093\\
& \cellcolor{cyan!20}\textbf{\name-DB}   &\cellcolor{cyan!20}\textbf{0.0388*}&\cellcolor{cyan!20}\textbf{0.0629*}&\cellcolor{cyan!20}\textbf{0.0254*}&\cellcolor{cyan!20}\textbf{0.0336*}&\cellcolor{cyan!20}\textbf{0.0868}&\cellcolor{cyan!20}\textbf{0.1074}&\cellcolor{cyan!20}\textbf{0.0747}&\cellcolor{cyan!20}\textbf{0.0807}&\cellcolor{cyan!20}\textbf{0.0221}&\cellcolor{cyan!20}\textbf{0.0345}&\cellcolor{cyan!20}\textbf{0.0145}&\cellcolor{cyan!20}\textbf{0.0185}\\
& \cellcolor{cyan!20}\textbf{\name-TB}   &\cellcolor{cyan!20}\textbf{0.0407*}&\cellcolor{cyan!20}\textbf{0.0651*}&\cellcolor{cyan!20}\textbf{0.0266*}&\cellcolor{cyan!20}\textbf{0.0344*}
&\cellcolor{cyan!20}\textbf{0.0882*}&\cellcolor{cyan!20}\textbf{0.1093*}&\cellcolor{cyan!20}\textbf{0.0747*}&\cellcolor{cyan!20}\textbf{0.0815*}  &\cellcolor{cyan!20}\textbf{0.0256*}&\cellcolor{cyan!20}\textbf{0.0404*}&\cellcolor{cyan!20}\textbf{0.0166*}&\cellcolor{cyan!20}\textbf{0.0213*}\\\midrule
\multirow{9}{*}{{LETTER}}       
& SFT  &    0.0344  &  0.0540     &  0.0227& 	0.0290    &   0.0853&0.1049&0.0734&0.0797  &   0.0205 	&    0.0341  &  0.0131  & 0.0175       \\
& GRPO      &  0.0384    &0.0592&0.0251&0.0317&
0.0819&0.0976&0.0709&0.0759&0.0228&0.0380&0.0146&0.0195\\
& DPO       &0.0358&0.0571&0.0238&0.0307&0.0788&0.0960&0.0682&0.0737&  0.0172  &    0.0286  &  0.0115   &   0.0151   \\
& S-DPO      &0.0390&0.0607&0.0259&0.0336&0.0769&0.0914&0.0674&0.0722&0.0247&0.0389&0.0161&0.0206\\
& SPRec      &0.0370&0.0573&0.0249&0.0315&0.0754&0.0896&0.0655&0.0701&0.0190&0.0303&0.0124&0.0160\\
& IPA      &0.0367&0.0586&0.0252&0.0320&0.0750&0.0891&0.0648&0.0699&0.0197&0.0293&0.0129&0.0163\\
& \cellcolor{cyan!20}\textbf{\name-DB}   & \cellcolor{cyan!20}\textbf{0.0407*}&\cellcolor{cyan!20}\textbf{0.0630*}&\cellcolor{cyan!20}\textbf{0.0266}&\cellcolor{cyan!20}\textbf{0.0337}&
\cellcolor{cyan!20}\textbf{0.0855}&\cellcolor{cyan!20}\textbf{0.1071 }&\cellcolor{cyan!20}\textbf{0.0738}&\cellcolor{cyan!20}\textbf{0.0802}
&\cellcolor{cyan!20}\textbf{0.0252}&\cellcolor{cyan!20}\textbf{0.0399}&\cellcolor{cyan!20}\textbf{0.0166}&\cellcolor{cyan!20}\textbf{0.0213}\\
&\cellcolor{cyan!20} \textbf{\name-TB}   &\cellcolor{cyan!20}\textbf{0.0433*}&\cellcolor{cyan!20}\textbf{0.0672*}&\cellcolor{cyan!20}\textbf{0.0286*}&\cellcolor{cyan!20}\textbf{0.0363*}&
\cellcolor{cyan!20}\textbf{0.0888*}&\cellcolor{cyan!20}\textbf{0.1100*}&\cellcolor{cyan!20}\textbf{0.0750*}&\cellcolor{cyan!20}\textbf{0.0819*}&
\cellcolor{cyan!20}\textbf{0.0261*}&\cellcolor{cyan!20}\textbf{0.0413*}& \cellcolor{cyan!20}\textbf{0.0173*}&\cellcolor{cyan!20}\textbf{0.0222*}               
\\\bottomrule
\end{tabular}%
\vspace{-5mm}
}
\end{table*}

\subsection{Experimental Setups}
\textbf{Datasets.}
In this paper, we conduct comprehensive experiments across three widely used datasets: Yelp, Beauty, and Instruments. Yelp\footnote{\url{https://www.yelp.com/dataset}} records user check-ins at restaurants, along with corresponding reviews and ratings. Beauty and Instruments are subcategories from Amazon Review\footnote{\url{https://cseweb.ucsd.edu/\~jmcauley/datasets.html\#amazon\_reviews}}~\cite{mcauley2015image}, which include user histories on Amazon products.  
For data preprocessing, we follow the procedures in LETTER~\cite{wang2024learnable}. Items are tokenized based on titles and descriptions for Beauty and Instruments, and based on titles, descriptions, brands, and categories for Yelp. For data splitting, we use the last interacted item for each user as the test sample, the second-to-last as the validation sample, and the remaining items as the training samples; i.e., each interacted data point is the target for the preceding user sequence. We present the data statistics in Table~\ref{tab:exp_dataset}. 

\noindent\textbf{Baselines and Backbones.}
We compare \name with three categories of LLM fine-tuning baselines: (a) vanilla \textbf{SFT}; (b) on-policy RLFT: \textbf{GRPO}~\cite{shao2024deepseekmathGRPO}; (c) off-policy RLFT: \textbf{DPO}~\cite{rafailov2023directDPO}, and DPO-based algorithms modified for language-based recommendations, including \textbf{S-DPO}~\cite{chensoftmax}, \textbf{SPRec}~\cite{gao2025sprec}, \textbf{IPA}~\cite{deng2025onerec}. Notably, the specially designed auxiliary loss, in addition to SFT, is considered part of the GR backbones. And we install \name to two backbones: \textbf{TIGER}~\cite{rajput2023recommender} and \textbf{LETTER}~\cite{wang2024learnable}. Detailed introduction is as follows:
\begin{itemize}[leftmargin=*]
\item \textbf{SFT}. Supervised fine-tuning, optimizes LLMs with the next token prediction loss as in Equation (1) of the main text. 
\item \textbf{GRPO}~\cite{shao2024deepseekmathGRPO}. Group Relative Policy Optimization is another on-policy RL algorithm based on the PPO idea. GRPO improves the PPO by updating the model based on the relative advantages of a set of actions, thereby reducing reliance on the absolute reward function. 
\item \textbf{DPO}~\cite{rafailov2023directDPO}. Direct Policy Optimization is an off-policy RL algorithm that updates the model using off-policy pairwise preference data, simplifying training while achieving competitive results.
\item \textbf{S-DPO}~\cite{chensoftmax}. 
This is a DPO-like algorithm tailored for language model-based recommendation scenarios. Unlike standard DPO, it can utilize multiple negative samples, making it better adapted to the characteristics of recommendation tasks.
\item \textbf{SPRec}~\cite{gao2025sprec}. This method selects rejected samples for DPO to mitigate biases at the item and token levels during LLM fine-tuning for recommendation tasks. Specifically, it updates rejected samples to negative samples based on the highest beam search scores and fine-tunes LLMs via iterative SFT and DPO.
\item \textbf{IPA}~\cite{deng2025onerec}. This method is part of OneRec, which aims to align LLMs with a reward model integrating collaborative knowledge. IPA iteratively updates the model via SFT and DPO steps, replacing rejected samples with the highest-reward samples at each step.
\end{itemize}
\noindent\textbf{Evaluation Metrics.}
We evaluate all methods using two metrics, namely \textit{Hit Rate (R)} and \textit{Normalized Discounted Cumulative Gain (NDCG)}, on top-5 and top-10 recommended lists, yielding four metrics per dataset: \textit{R@5}, \textit{R@10}, \textit{N@5}, and \textit{N@10}. To ensure robust results, each test is repeated three times with different random seeds (\{42, 43, 44\}).

\noindent\textbf{Implementation Details.}
We run all experiments on the same GPU. Regarding the hyperparameter settings, we find the optimal settings for both baselines and \name. The default ensemble of \name takes the CM-based sampling for the trajectory sampler, and the sum of three reward signals as the reward, i.e., $\mathcal{R}(\tau)=r_a+r_c+r_{\text{sim}}$.
The training procedure is controlled by Transformer Trainers with early stopping. The implementation of baselines is based on TRL (GRPO, DPO) and the released code of papers (S-DPO, SPRec). We implement IPA ourselves, using the same collaborative reward model as for \name. Code is available at \url{https://github.com/Applied-Machine-Learning-Lab/SIGIR26_GFlowGR}.


\subsection{Main Results (RQ1)}
To evaluate the effectiveness of \name, we tested its two variants (DB and TB) against various fine-tuning methods. In this experiment, we instantiate \name with CM-based sampling in Section~\ref{sec:sampl} and the sum of three reward signals in Section~\ref{sec:rew}. The results are summarized in Table~\ref{tab:ovall}, which consistently demonstrates that \name outperforms baselines. Detailed observations include:

\noindent\textbf{Comparison with SFT. }
Both \name variants consistently outperform SFT across all datasets and metrics, demonstrating that \name effectively enhances model capacity through augmented samples, flow-matching objectives, and integrated collaborative knowledge. Notably, \name-TB achieves more significant improvements than \name-DB, particularly on Yelp with TIGER, likely because GR’s intermediate states do not uniquely identify items, making whole-trajectory optimization (via TB’s trajectory-level balance) more effective by leveraging one-to-one trajectory-item correspondences.

\noindent\textbf{Comparison with On-policy RLFT. }
On-policy strategies fail to achieve competitive results for GR. This may stem from their sensitivity to the data. Although GRPO can also learn from an item set with relative advantages.~\cite{shao2024deepseekmathGRPO}, it maximizes the probability for the positive item, failing to concept the out-of-set items. While \name assigns the probability proportional to the reward without forcing the sum of the item set as 1, it possesses better robustness in terms of the augmented item set. 

\noindent\textbf{Comparison with Off-policy RLFT. }
Vanilla DPO~\cite{rafailov2023directDPO} fails to stably improve upon SFT, particularly on large-scale Yelp data, due to its use of fixed and randomly sampled ``Rejected'' items. In contrast, SPRec~\cite{gao2025sprec} (using self-play sampling) and IPA~\cite{deng2025onerec} (leveraging collaborative knowledge) enhance DPO through dynamic, guided negative sampling. Among DPO-based methods, S-DPO~\cite{chensoftmax}, which can incorporate multiple negative samples, achieves the best results and closely competes with \name-DB. This also highlights the urgent need for the ability of learning from the item set rather than a single positive item or item pairs.

\begin{figure*}[t]
\centering
\includegraphics[width=1\textwidth]{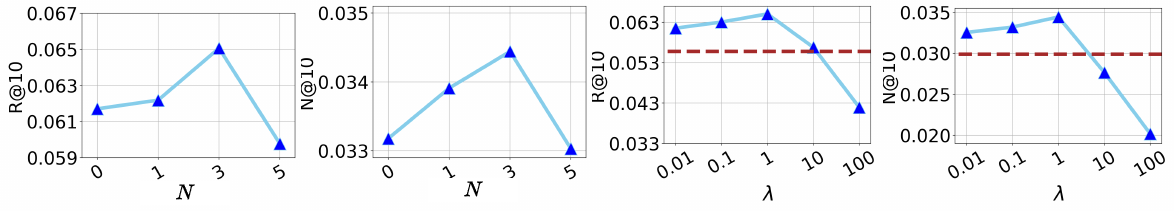}
\vspace{-7mm}
\caption{Parameter study on $N$ and $\lambda$ with \textbf{TIGER} on \textbf{Beauty}. The red dashed line denotes the performance of SFT, which falls outside the plotted range for the figures corresponding to $N$.}
\label{fig:param1}
\vspace{-2mm}
\end{figure*}

\begin{figure*}[t]
\centering
\vspace{-2mm}
\includegraphics[width=1\textwidth]{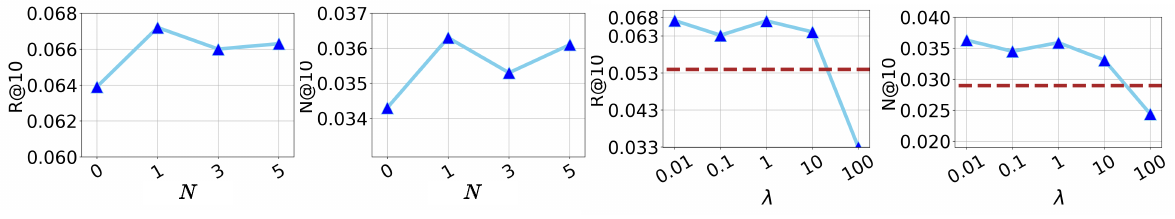}
\vspace{-7mm}
\caption{Parameter study on $N$ and $\lambda$ with \textbf{LETTER} on \textbf{Beauty}. The red dashed line denotes the performance of SFT, which falls outside the plotted range for the figures corresponding to $N$.}
\label{fig:param2}
\vspace{-2mm}
\end{figure*}

\subsection{Parameter Analysis (RQ2)}
In this section, we analyze \name’s parameter sensitivity, focusing on the number of augmented trajectories $K \in \{0, 1, 3, 5\}$ and the balancing weight $\lambda \in \{0.01, 0.1, 1, 10, 100\}$ in Equation~\eqref{eq:final_loss}. Figure~\ref{fig:param1} and Figure~\ref{fig:param2} visualize the evaluation results for TIGER and LETTER on Beauty, respectively. As $K$ and $\lambda$ increase, performance follows a reverse `V' curve: \textbf{For $K$}, small values (e.g., $K=0$) fail to provide sufficient diversity for sparse data, while large $K=5$ introduces noisy augmented samples that mislead training.\textbf{ For $\lambda$}, a moderate weight of $1.0$ optimizes $\mathcal{L}_{\text{GFN}}$, balancing flow-matching objectives with the core recommendation task. Overemphasizing $\mathcal{L}_{\text{GFN}}$ (e.g., $\lambda \geq 10$) degrades GR performance, highlighting the critical role of SFT in maintaining basic recommendation capabilities.  


\begin{table*}[t]
\tabcolsep=0.13cm 
\centering
\caption{Ablation study with LETTER on Beauty.}
\label{tab:abl}
\begin{tabular}{@{}c|c|cccccc|ccc|c@{}}
\toprule
Model&Metrics& $r_a$ & $r_c$ & $r_{\text{sim}}$  & $r_a+r_c$ & $r_a+r_{\text{sim}}$  & $r_c+r_{\text{sim}}$ & \textit{w/o} Ada & \textit{w/o} Traj & \textit{w/o} CM&\cellcolor{cyan!20}\textbf{\name}  \\ \midrule
\multirow{4}{*}{LETTER}  &R@5&0.0387&0.0408&0.0384 &0.0398&0.0404&0.0410&0.0424&0.0423&0.0430&\cellcolor{cyan!20}\textbf{0.0433}\\
&R@10&0.0596&0.0631&0.0580&0.0610&0.0603&0.0622&0.0659&0.0639&0.0667&\cellcolor{cyan!20}\textbf{0.0672}\\
&N@5&0.0254&0.0265&0.0246&0.0259&0.0263&0.0274&0.0278&0.0274&0.0282&\cellcolor{cyan!20}\textbf{0.0286}\\
&N@10&0.0322&0.0337&0.0310&0.0327&0.0327&0.0342&0.0354&0.0343&0.0358&\cellcolor{cyan!20}\textbf{0.0363}\\\midrule
\bottomrule
\end{tabular}%
\end{table*}

\subsection{Ablation Study (RQ3)}\label{sec:abl}
In this section, we validate the critical components of \name: 
The trajectory sampler and the reward model. 
Table~\ref{tab:abl} presents ablation results for: all reward signal combinations (six variants, first block of the table, where the name of the variant denotes used reward signals), removal of CM-based adaptive sampling (w/o Ada, i.e., using random sampling), omission of Augmented Trajectories (w/o Traj), i.e., $N=1$. 
To assess the reliance on CM, we evaluate a ``w/o CM'' variant that eliminates CM entirely: removing its outputs from both the reward model and the trajectory construction. Key findings include: (a) \textbf{Loss of any reward signal} (augmentation signals, collaborative scores, or token similarities) degrades performance, confirming the necessity of the comprehensive reward model. 
Notably, $r_a$ has the greatest impact, while $r_c$ contributes marginally because CM produces less accurate predictions than GR~\cite{rajput2023recommender,wang2024learnable}, leaving noisy $r_c$ signals.
(b) \textbf{Using random sampling} reduces effectiveness due to unstable training or low-quality augmented samples.  
(c) \textbf{Learning from the single positive} leads to worse generalization due to the lack of diverse training trajectories, which limits the model’s ability to explore high-reward item regions and adapt to the set-wise generation demands of real-world deployment. 
(d) \textbf{\name demonstrates robustness in a CM-free setting}, retaining strong performance even without collaborative signals. This indicates that its value-aware, set-wise design can effectively guide generation even without CM.


\begin{table}[t]
\caption{Comparison of reward strategies on Beauty.}
\label{tab:reward_fusion}
\centering
\begin{tabular}{lcccc}
\toprule
\textbf{Method} & \textbf{R@5} & \textbf{R@10} & \textbf{N@5} & \textbf{N@10}\\
\midrule
Deep Integration & 0.0399 & 0.0622 & 0.0263 & 0.0334\\
Weighted Sum & \textbf{0.0446} & \textbf{0.0673} & \textbf{0.0293} & \textbf{0.0366}\\
Sum (default) & 0.0433 & 0.0672 & 0.0286 & 0.0363\\
\bottomrule
\end{tabular}
\end{table}

\subsection{Reward Integration Analysis (RQ4)}
Beyond the fixed summation strategy used in our main experiments, we also explore adaptive fusion of the three reward signals. Specifically, we compare three integration approaches: (1) \textit{Sum}: simple addition (our default); (2) \textit{Deep Integration}: a learnable fusion via a multi-layer perceptron; and (3) \textit{Weighted Sum}: a lightweight learnable linear combination with trainable scalar weights per signal. Table~\ref{tab:reward_fusion} reports results on Beauty using LETTER.

The learnable \textit{Weighted Sum} consistently improves performance at top-5 metrics (R@5 and N@5), while matching the strong results of \textit{Sum} at top-10. This suggests that even a minimal learnable mechanism can better calibrate the relative importance of heterogeneous signals. While the over-complex \textit{Deep Integration} fails to provide the enhancement. These findings confirm that \name's performance can be further enhanced through refined reward design, even though the default summation already yields strong results.

\subsection{LLM-enhancement Study (RQ5)}
We explore two lightweight ways to enhance \name using LLMs in GR: sampling and rewards.

\noindent\textbf{LLM-based Sampling.}
Instead of sampling augmented trajectories randomly or solely via CM, we use the LLM’s generation confidence to select candidates from a CM-filtered pool. As Table~\ref{tab:sample_selection} shows, both max and min confidence strategies improve over the default (last row), with low-confidence selection boosting R@5/N@5—suggesting LLM-guided exploration enhances diversity without degrading relevance.

\noindent\textbf{LLM-based Reward.}
We add the LLM’s normalized generation probability of an item as an auxiliary reward signal—requiring no extra inference, as it is already computed during decoding. Table~\ref{tab:llm_reward} shows consistent gains in R@5 and N@5, indicating that LLM probabilities provide valuable value-aware supervision beyond explicit interactions.
Together, these results show that even minimal integration of LLM signals—during both trajectory construction and reward estimation—can meaningfully improve \name.

\begin{table}[t]
\caption{Performance comparison of different sample selection strategies on Beauty (LETTER).}
\label{tab:sample_selection}
\centering
\begin{tabular}{lccccc}
\toprule
Method & Selection & R@5 & R@10 & N@5 & N@10\\
\midrule
Min & 1 of 2 & 0.0442 & 0.0654 & 0.0293 & 0.0361\\
Max & 1 of 2 & 0.0432 & 0.0650 & 0.0285 & 0.0355\\
Max & 1 of 5 & 0.0436 & 0.0650 & 0.0287 & 0.0355\\
GFlowGR & 1 of 1 & 0.0433 & 0.0672 & 0.0286 & 0.0363\\
\bottomrule
\end{tabular}
\vspace{-3mm}
\end{table}
 
\begin{table}[t]
\caption{Performance comparison with and without LLM-based reward on Beauty (LETTER).}
\label{tab:llm_reward}
\centering
\begin{tabular}{lcccc}
\toprule
Method & R@5 & R@10 & N@5 & N@10 \\
\midrule
with LLM reward & \textbf{0.0458} & 0.0670 & \textbf{0.0301} & 0.0368\\
without LLM reward  & 0.0433 & \textbf{0.0672} & 0.0286 & \textbf{0.0363}\\
\bottomrule
\end{tabular}
\vspace{-3mm}
\end{table}

\begin{figure*}[t]
\centering
\includegraphics[width=0.8\textwidth]{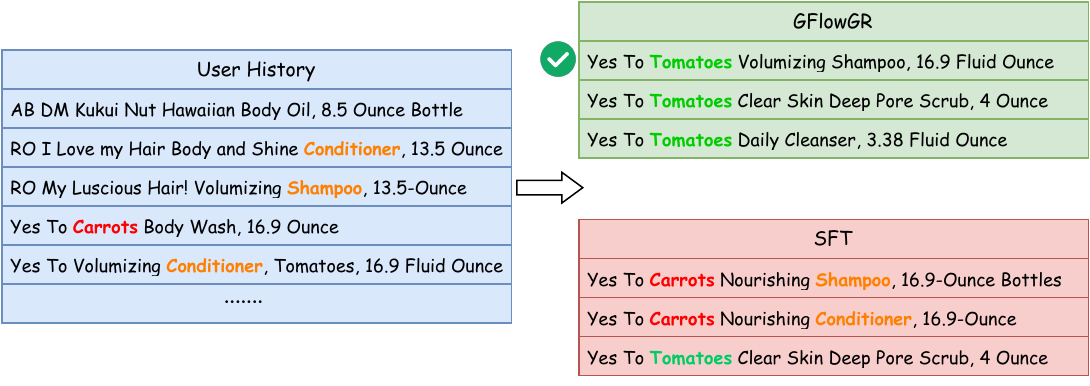}
\caption{\textbf{Case study on user-2089 from Beauty with TIGER}: We visualized three items from the top-10 list of each method for the brand `Yes to', where the top-1 result of \name is correct.}
\label{fig:case}
\vspace{-4mm}
\end{figure*}

\subsection{In-depth Analysis(RQ6)}
This section explores the diversity of \name and provides a case study with TIGER on Beauty. 


\noindent\textbf{Diversity.} 
For this experiment, we collect the top-10 recommendation list from \name and SFT and analyze the diversity from two perspectives: (a) Generation distribution over all items. We compute the probability of being recommended by the model for all items in the whole test data. Considering the intrinsic popularity difference of items, we compare this generation distribution with the normal distribution by KL Divergence~\cite{kullback1951kullback}. The result shows that \name can generate items with a probability distribution closer to normal than SFT models (KL divergence: $3.13<3.24$). In addition, \name has a smoother distribution, with a 2.2\% smaller standard deviation, indicating better diversity. (b) Collaborative scores for the recommended items. We calculate the collaborative score for all items within the collected list. We find that \name has a smaller average score but a broader range (the maximum minus the minimum, 0.45 for \name and 0.42 for SFT) and a 3.6\% larger standard deviation. This shows that, compared with SFT, \name is more willing to recommend items that are not regarded as competitive by the collaborative model (lower means) and to generate more diverse choices (larger range and greater deviation).  

\noindent\textbf{Case Study.}
Figure~\ref{fig:case} illustrates how \name generates recommendations that are both diverse and value-aligned. While SFT repeats ``Carrots''-flavored shampoo and conditioner (bold orange/red), which are items from the user’s recent history, \name recommends a varied set including ``Shampoo'', ``Scrub'', and ``Cleanser'' with the novel ``Tomatoes'' flavor (bold green). This shift arises because \name trains on sets of candidate trajectories and identifies high-utility combinations beyond surface-level interactions. As a result, it achieves greater diversity and successfully predicts the user's preferred item.


\begin{table}[t]
\caption{Offline evaluations on production data.}
\label{tab:off}
\begin{tabular}{@{}c|cccc@{}}
\toprule
\textbf{Method} & \textbf{H@20} & \textbf{N@20} & \textbf{Samples}&\textbf{Cost} \\ \midrule
SFT             & 0.358       & 0.371         &1$\times$&1.0$\times$                   \\
GRPO            & 0.427       & 0.408         &80$\times$&-              \\
\name & 0.444      & 0.452         &80$\times$&2.1$\times$              \\ \bottomrule
\end{tabular}
\vspace{-6mm}
\end{table}

\section{Deployment and Online Evaluations}

\begin{figure}[t]
\centering
\includegraphics[width=0.9\linewidth]{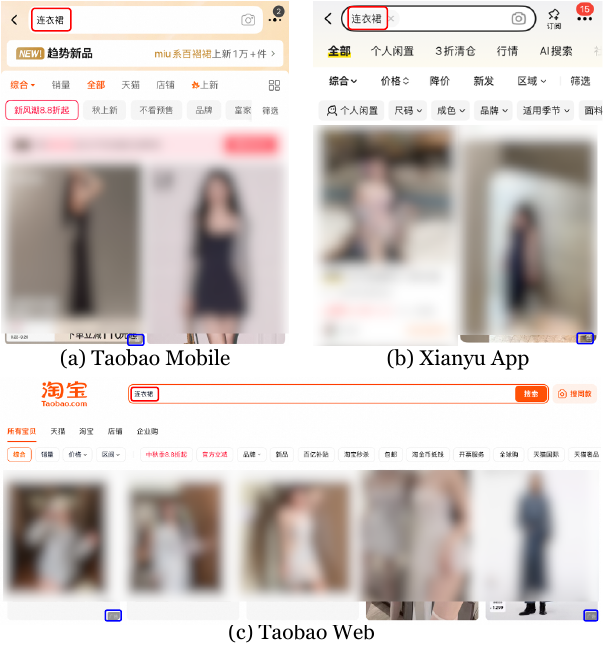}
\caption{Deployed business scenarios in Taobao, illustrated with the query ``Dress'' in the top left red box. Advertisements are indicated by blue boxes on the bottom right. Sensitive information has been obfuscated for privacy.}
\label{fig:dpl}
\end{figure}

Since May 2025, \name has been fully deployed to train our GR models across all Taobao search advertising platforms, including Taobao Mobile, Xianyu, and Taobao Web (as illustrated in Figure~\ref{fig:dpl}).

Offline evaluations, summarized in Table~\ref{tab:off}, compare \name against GRPO and SFT on a massive dataset of hundreds of millions of daily requests. The results show that \name achieves \textbf{H@20 of 0.444 and N@20 of 0.452}, substantially outperforming \textbf{SFT (0.358/0.371)} and \textbf{GRPO (0.427/0.408)}, mirroring the conclusions on public datasets. To maintain confidentiality, training costs are presented relative to the SFT baseline. While GRPO's cost is computationally infeasible, \name requires approximately double the training time of SFT. 
Despite using the same \textbf{80$\times$} augmented sample budget as GRPO, \name incurs only \textbf{2.1$\times$} the training cost of single-sample SFT. This corresponds to approximately \textbf{40$\times$} higher efficiency in user interaction utilization, and the remaining compute overhead can be eliminated through linear GPU scaling.


A 15-day online A/B test on 10\% of traffic (50/50 split) revealed a statistically significant \textbf{+0.43\%} lift in total revenue. A 1.1\% cost increment was observed in the low-latency-sensitivity off-home-page setting. Specifically, Taobao Mobile saw \textbf{+0.95\%} ($p<0.001,\sigma=0.1$) cost efficiency and \textbf{+0.69\%} CTR lift, while other segments (small scenarios like Xianyu and Taobao Web) achieved \textbf{+2.90\%} ($p<0.001,\sigma=0.5$) and \textbf{+0.39\%}, respectively. 
Notably, \name improves performance on under-served cases: \textbf{+4.0\%} revenue on newly listed items and \textbf{+1.8\%} on long-tail queries. The diversity also increases by \textbf{+0.11\%} in item coverage, confirming broader recommendations.

\section{Related Work}
\textbf{Generative Recommendations.} 
According to the GR composition, existing research can be generally divided into three categories: pipeline design, innovations in the item tokenizer, and LLM decoding enhancement. Representative works in the first category include TIGER~\cite{rajput2023recommender}, LC-Rec~\cite{zheng2024adapting}, and ColaRec~\cite{wang2024content}. TIGER~\cite{rajput2023recommender} is the pioneer in exploring expressing items with out-of-vocabulary tokens and constructs the basic GR pipeline, including the training and inference with item tokenizer and LLMs. LC-Rec~\cite{zheng2024adapting} further enhances the model's efficacy by incorporating diverse alignment tasks to align item identifiers with recommendation objectives. Similarly, ColaRec~\cite{wang2024content} introduces the graph-based collaborative models to align the semantics with recommendations. Some researchers also explore GR frameworks for other recommendation problems, such as multi-behavior recommendations~\cite{liu2024multi}. The second category focuses on inventing plug-in item tokenizers to enhance the performance of arbitrary GR frameworks~\cite{zhu2024cost,zheng2025universal,liu2024end,lin2025order}. For example, LETTER~\cite{wang2024learnable} introduces the collaborative item embeddings and diversity constraints to control the item tokenization process. While the last category focuses on the decoding process for GR and designs decoders for specific recommendation targets~\cite{lin2025efficient,ding2024inductive}. While few efforts have been devoted to LLM fine-tuning specific to GR beyond optimization algorithms for general language-based recommendations~\cite{chensoftmax,gao2025sprec}. IPA~\cite{deng2025onerec} and PROMISE~\cite{guo2026promise} are pioneers exploring this direction. The proposed \name is the first to establish the objective of value-aware set-wise learning.

\noindent\textbf{Generative Flow Network Applications.} 
GFlowNets are designed for compositional generation to produce diverse, high-reward outputs~\cite{bengio2021flow,bengio2023gflownet}, with applications spanning biology~\cite{jain2022biological}, scientific discovery~\cite{jain2023gflownets}, and multi-step reasoning~\cite{yu2024flowfor,kang2025gflowvlm}. While prior work has applied GFlowNets to discriminative recommendation tasks like sequential recommendation~\cite{liu2024modeling} and list-wise recommendation~\cite{liu2023generative}, as well as language-based recommendations~\cite{gao2025process}, we are the first to explore GFlowNets for generative recommendations.

\section{Conclusion}

In this work, we propose \name, a novel generative recommendation framework that leverages Generative Flow Networks to bridge the gap between point-wise training and set-wise serving. By modeling item generation as a sequential trajectory and aligning generation probabilities with behavior-aware rewards, \name enables value-proportional, diverse, and high-quality recommendations. We design a comprehensive reward model that fuses multiple signals and introduce an adaptive trajectory sampler that efficiently explores high-utility candidates. Extensive experiments on public and large-scale production datasets demonstrate that \name consistently outperforms strong baselines such as SFT and GRPO, both offline and in real-world deployment. Notably, online A/B tests involving over 100 million user sessions show significant improvements in revenue. Our results validate that GFlowNets provide a principled and practical foundation for next-generation generative recommender systems.
\begin{acks}
This research was partially supported by National Natural Science Foundation of China (No.62502404), Hong Kong Research Grants Council (Research Impact Fund No.R1015-23, Collaborative Research Fund No.C1043-24GF, General Research Fund No. 11218325), Institute of Digital Medicine of City University of Hong Kong (No.9229503), and CCF-Alimama Tech Kangaroo Fund No. 2024002.
\end{acks}

\clearpage
\bibliographystyle{ACM-Reference-Format}
\balance
\bibliography{main}


\end{sloppypar}
\end{document}